\documentclass[prl, twocolumn,superscriptaddress,preprintnumbers,a4paper,amsmath,amssymb,showpacs,floatfix]{revtex4}

\usepackage{graphicx}
\usepackage{color}\color[rgb]{0.000,0.000,0.000} 
\usepackage{amssymb} 
\usepackage{amsmath} 
\usepackage[normalem]{ulem} 
\usepackage {ulem} 

\begin{document}

\newcommand{\pr}{PrMnSbO}
\newcommand{\prf}{PrFeAsO$_{1-x}$F$_{x}$}
\newcommand{\prsc}{PrFeAsO$_{0.85}$F$_{0.15}$}

\newcommand{\te}{$_{t}$}
\newcommand{\ot}{$_{o}$}
\newcommand{\Tc}{$T_\mathrm{C}$}
\newcommand{\Ts}{$T_\mathrm{s}$}
\newcommand{\Tn}{$T_\mathrm{N}$}
\newcommand{\Tnpr}{$T_\mathrm{sN}$(Pr)}
\newcommand{\Tnfe}{$T_\mathrm{N}$(Fe)}

\newcommand{\Mueff}{$\mu_{\mathrm{eff}}$}
\newcommand{\MuB}{$\mu_\mathrm{B}$}


\title{Local moments and symmetry breaking in metallic \pr}

\author{Simon A.J. Kimber}
\email[Email of corresponding author:]{kimber@esrf.fr}

\affiliation{European Synchrotron Radiation Facility (ESRF), 6 rue Jules Horowitz, BP 220, 38043  Grenoble Cedex 9, France}
\affiliation{Helmholtz-Zentrum Berlin f\"ur Materialien und Energie (HZB), Glienicker Strasse 100, D-14109, Berlin, Germany}

\author{Adrian H. Hill}

\affiliation{European Synchrotron Radiation Facility (ESRF), 6 rue Jules Horowitz, BP 220, 38043  Grenoble Cedex 9, France}

\author{Yu-Zhong Zhang}

\affiliation{Institut f\"ur Theoretische Physik, Goethe-Universit\"at Frankfurt, Max-von-Laue-Stra\ss e 1, 60438 Frankfurt am Main, Germany}

\author{Harald O. Jeschke}

\affiliation{Institut f\"ur Theoretische Physik, Goethe-Universit\"at Frankfurt, Max-von-Laue-Stra\ss e 1, 60438 Frankfurt am Main, Germany}

\author{Roser Valent\'{i}}

\affiliation{Institut f\"ur Theoretische Physik, Goethe-Universit\"at Frankfurt, Max-von-Laue-Stra\ss e 1, 60438 Frankfurt am Main, Germany}

\author{Clemens Ritter}

\affiliation{Institute Max von Laue-Paul Langevin, 6 rue Jules Horowitz, BP 156, F-38042, Grenoble Cedex 9, France.}

\author{Inga Schellenberg}

\affiliation{Westf\"alische Wilhelms-Universit\"at M\"unster, Corrensstrasse 30, 48149 M\"unster, Germany}

\author{Wilfred Hermes}

\affiliation{Westf\"alische Wilhelms-Universit\"at M\"unster, Corrensstrasse 30, 48149 M\"unster, Germany}

\author{Rainer P\"ottgen}

\affiliation{Westf\"alische Wilhelms-Universit\"at M\"unster, Corrensstrasse 30, 48149 M\"unster, Germany}

\author{Dimitri N. Argyriou}
\email[Email of corresponding author:]{argyriou@helmholtz-berlin.de}
\affiliation{Helmholtz-Zentrum Berlin f\"ur Materialien und Energie (HZB), Glienicker Strasse 100, D-14109, Berlin, Germany}

\date{\today}

\pacs{75.50.-y,75.25.-j,71.20.}
\begin{abstract}
We report a combined experimental and theoretical investigation of the layered antimonide \pr \  which is isostructural to the parent phase of the iron pnictide superconductors. We find linear resistivity near room temperature and Fermi liquid-like $T^{2}$ behaviour below 150 K. Neutron powder diffraction shows that unfrustrated C-type Mn magnetic order develops below $\sim$ 230 K, followed by a spin-flop coupled to induced Pr order. At T $\sim$ 35 K, we find a tetragonal to orthorhombic  (T-O) transition. First principles calculations show that the large magnetic moments observed in this metallic compound are of local origin. Our results are thus inconsistent with either the itinerant or frustrated models proposed for symmetry breaking in the iron pnictides. We show that \pr \ is instead a rare example of a metal where structural distortions are driven by $f$-electron degrees of freedom.
\end{abstract}

\maketitle
The discovery \cite{kamihara} of high temperature superconductivity in LaFeAsO$_{1-x}$F$_{x}$ has re-ignited interest in the role of magnetism in this phenomenon. The characteristic feature of the parent, non-superconducting, iron pnictides is a  tetragonal- orthorhombic (T-O) structural phase transition, which accompanies or precedes striped magnetic order \ \cite{dope1}. Optimal \Tc \ is reached after this transition is completely suppressed by alo- or isovalent doping \cite{dope1,dope}, or by the application of pressure \cite{me}. Despite metallic electrical conduction and even reports of quantum oscillations \cite{QO}, the origins of the magnetism in the iron pnicitdes are currently controversial \cite{theory}. The key experimental observations of a strongly reduced magnetic moment and symmetry breaking of the square FeAs substructure, can to some degree, be explained by two competing viewpoints. The first, is based on itinerant models of magnetism \cite{yz}, while the second is based upon local moment exchange as the iron lattice maps onto a frustrated $J_{1}-J_{2}$ model \cite{Yildirim}. Currently consensus seems far away, but what is not disputed is that magnetism, structural distortion, and superconductivity are related. Understanding the correlations between structure, magnetism and transport properties in the whole family of oxypnictide materials \cite{zrcusias} \ is thus of great topical interest.
\\In this article we report a combined experimental and theoretical investigation of \pr, which is isostructural to the LaFeAsO parent phase of the iron pnictide superconductors \cite{orig,synth}. We find strong evidence for localized magnetic moments which co-exist with a sea of conduction electrons, and a T-O structural phase transition at 35 K. Even though Mn spin order is found at extremely high temperature (\Tn \ = 230 K), this barely couples to the lattice showing that neither the itinerant, nor frustrated models for the iron arsenides are appropriate. Instead, the symmetry breaking transition is apparently driven by single ion Pr$^{3+}$ degrees of freedom. While many other isostructural materials with $f$-electron degrees of freedom are known \cite{sm,hf,jan,ndco}, none have previously been shown to undergo structural phase transitions except the iron pnictide parent phases.
\begin{figure}[tb!]
\begin{center}
\includegraphics[scale=0.35]{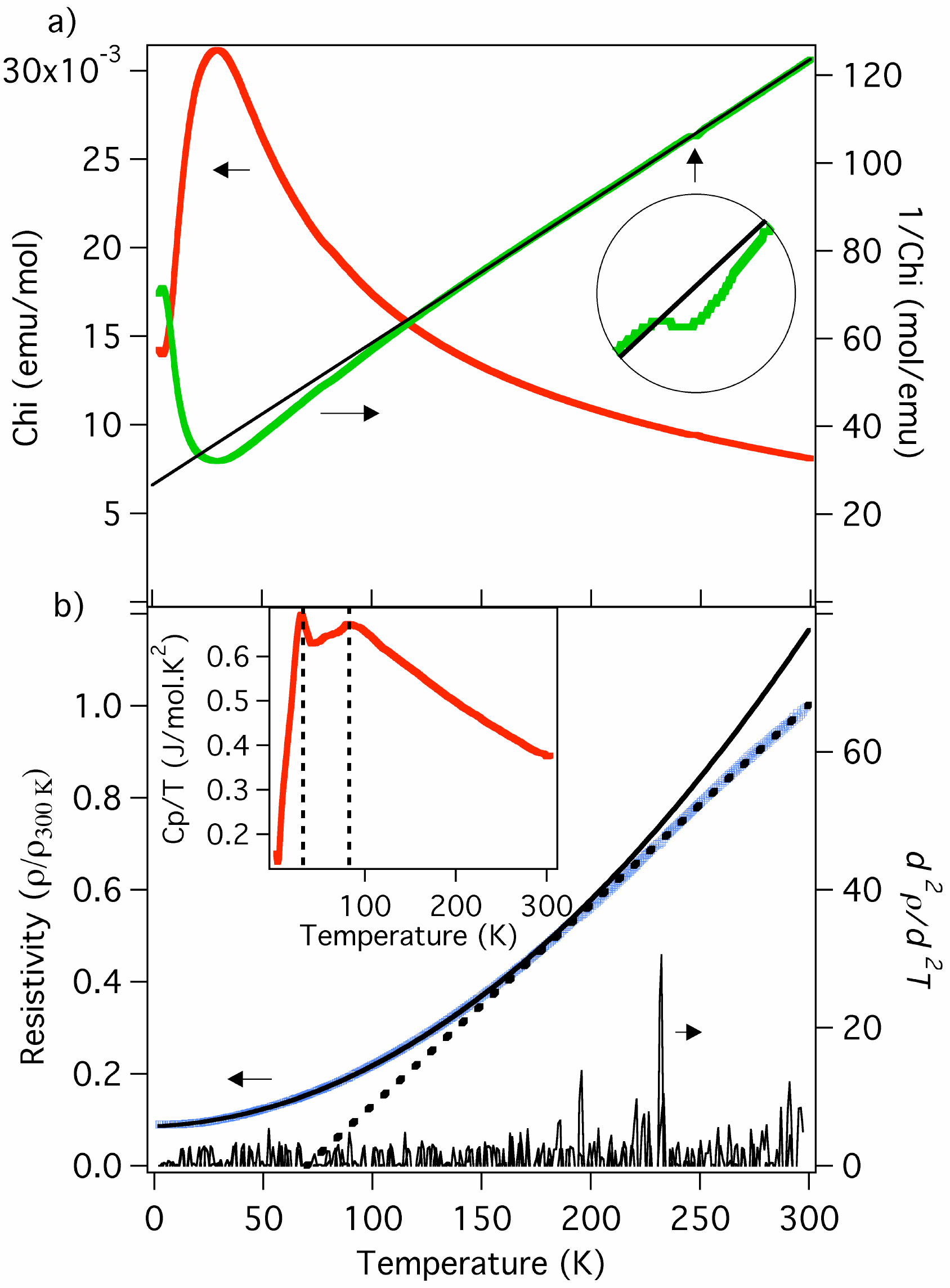}
\caption{(color online) (a) DC-magnetic susceptibility of \pr\ measured as a function of temperature in a 1 T field; (b) Resistivity of polycrystalline \pr\ measured in zero field, solid line shows fit to $T^{2}$ behaviour over range 2 $< \ T \ <$ 150 K, dashed line shows linear behaviour near room temperature.}
\label{fig1}
\end{center}
\end{figure}
\\ We synthesised polycrystalline samples of \pr \ using the previously reported method \cite{synth} and measured a range of physical properties. Magnetic susceptibility measurements (Fig. 1a), performed with a Quantum Design MPMS system in a 1 T field, show an anomaly near $\sim$ 250 K and an antiferromagnetic transition at around 40 K. The inverse susceptibility was well fitted by a Curie-Weiss term over a wide temperature region, and gave an effective moment of 4.99(2)   \MuB \  and a Weiss constant of -85(1) K, implying strong antiferromagnetic correlations. Assuming Pr$^{3+}$ has the expected free ion moment (in agreement with the neutron diffraction results below) of 3.58  \MuB \ , then the paramagnetic Mn moment is 3.47  \MuB. This value is consistent with the very strong Mn-Sb hybridisation in this class of materials which stabilises an intermediate spin configuration \cite{bs}.
\\ Transport measurements (Fig. 1b) were performed on \pr\ using a PPMS system, and showed a low room temperature resistivity of 0.011 $\Omega\cdot$m. The temperature dependance was metallic, with a residual resistivity of $\rho/\rho_{300\ K}$ = 0.09. Fermi liquid-like $T^{2}$ behaviour was found over a wide temperature regime (2 - 150 K) with a crossover at higher temperatures to quasi-linear behaviour.  An anomaly in $d^{2}\rho/d^{2}T$ was found at 230 K (Fig. 1b). As shown below, this corresponds to the Mn ordering temperature. The linear resistivity at higher temperatures could therefore be a result of Mn spin fluctuations. Intriguingly, the succession of magnetic and structural transitions which  occur at lower temperatures, have no further effect on the transport properties. These are however, clearly visible in the specific heat (inset Fig. 1b), which has a sharp peak at 35 K and a broader feature centered at 80 K. The 230 K transition is not clearly distinguished, probably due to the very large phonon contribution to C$_{p}$(T) at high temperature.
\begin{figure}[tb!]
\begin{center}
\includegraphics[scale=0.4]{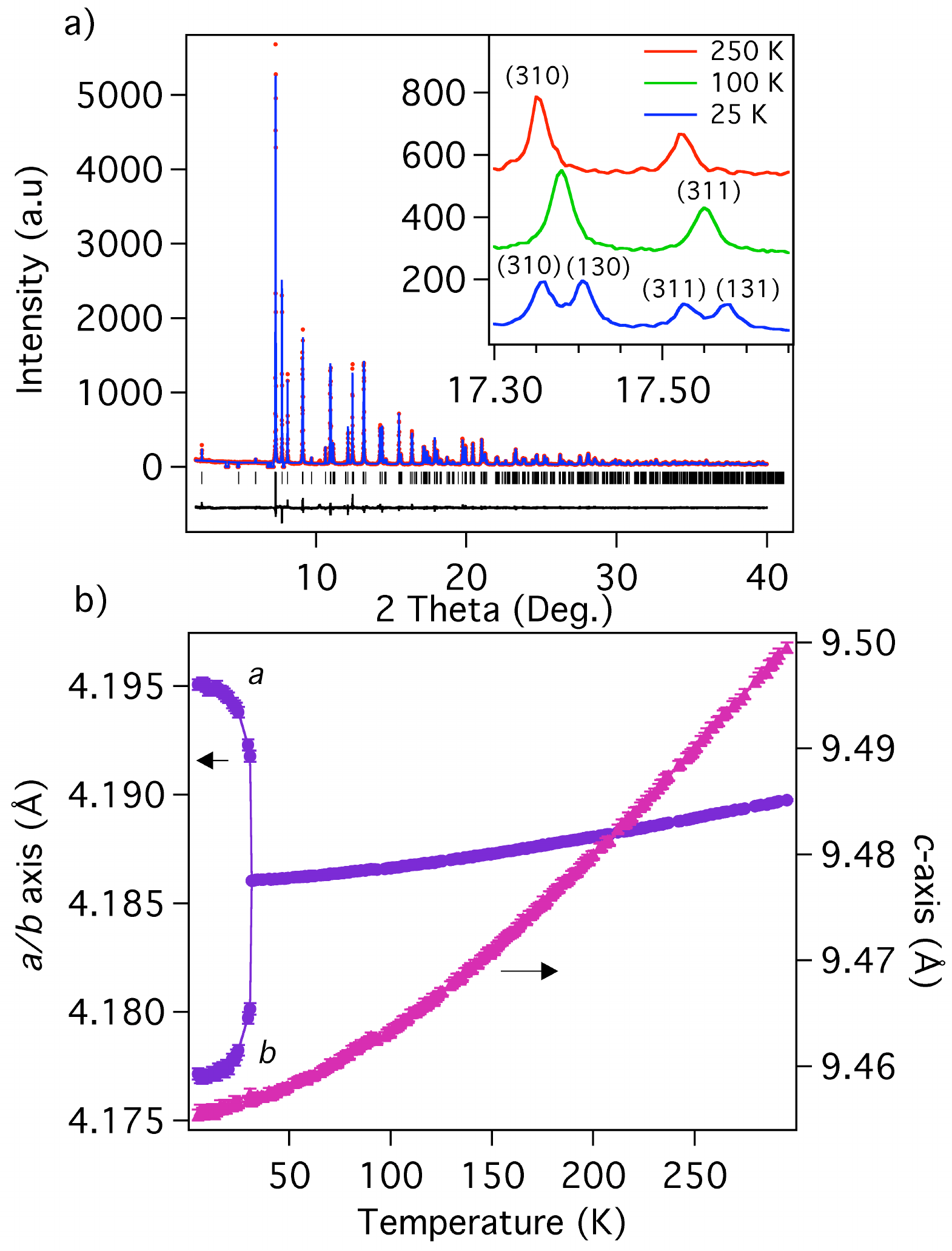}
\caption{(color online) (a) Observed, calculated and difference profiles for Rietveld fit to synchrotron X-ray diffraction profile of \pr\ at 20 K. The refinement converged with $R_{wp}$ = 0.067, $R_{p}$ = 0.052 and $\chi^{2}$ = 2.89. The inset shows the orthorhombic peak splitting as a function of temperature. (b) Refined lattice parameters of \pr\  as a function of temperature from the neutron powder diffraction experiment.}
\label{fig2}
\end{center}
\end{figure}
\\At room temperature, Rietveld analysis \cite{GSAS} of extremely high resolution X-ray powder diffraction data collected using ID31 at the ESRF, Grenoble, were consistent with the reported \ \cite{synth} \  $P4/nmm$ structure and no evidence for disorder or deficiencies was detected. Below $\sim$35 K, a clear orthorhombic splitting is seen (inset, Fig. 2a). Unlike LaFeAsO, this does not lead to an enlarged centred cell. Here the structure was solved and refined in the subgroup $Pmmn$ with a small distortion in the $ab$ plane \cite{coord}. No change in the nearest neighbour Mn-Mn distance was found, however, the Pr site point group symmetry is reduced from $4mm$ to $2m$. This is consistent with the rare earth magnetism driving the structural distortion as confirmed below. The temperature dependence of the cell parameters is shown in Fig. 2b. Clearly, the manganese spin ordering transition at 230 K has little or no coupling to the lattice as no anomalies are seen. We note that the absolute magnitude of the distortion in the $ab$ plane at 35 K (4.2 \%)  is only slightly smaller than that found \cite{prfeaso} \  in PrFeAsO (4.8 \%), although the latter undergoes the T-O phase transition at significantly higher temperatures (136 K)
\begin{figure*}[tb!]
\begin{center}
\includegraphics[scale=0.95]{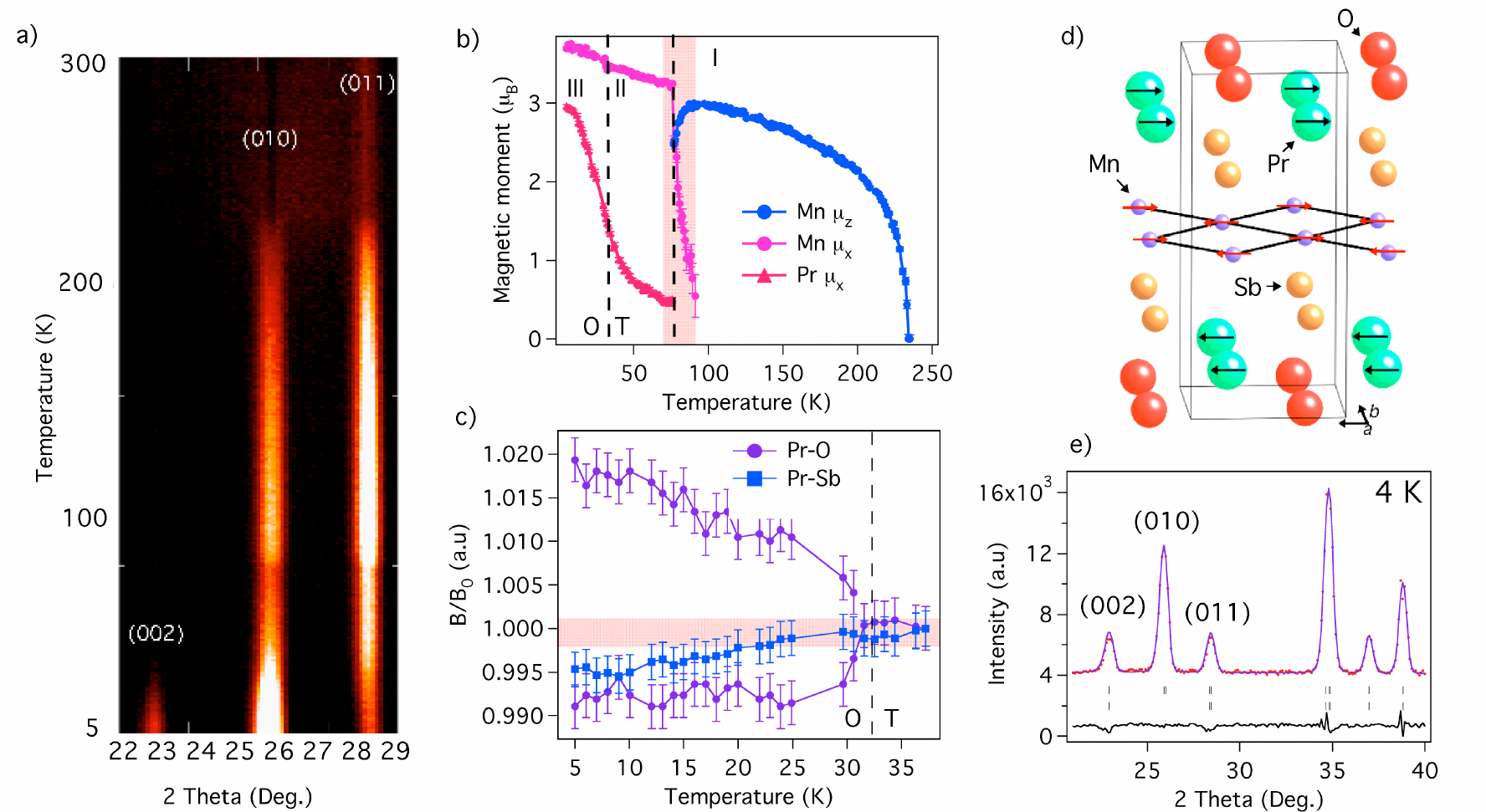}
\caption{(color online) (a) Temperature dependence of principle magnetic reflections in \pr\ from neutron powder diffraction, labels refer to the orthorhombic cell; (b) Refined magnetic moments through regimes referred to in the text as a function of temperature, the region over which the spin-flop occurs is indicated in pink; (c) Relative changes of Pr-O and Pr-Sb bond distances below the T-O phase transition, the pink area indicates the maximum changes in these distances found in PrFeAsO; (d) Magnetic structure of \pr\ obtained at 4 K; (e) Low angle part of observed, calculated and difference profiles for Rietveld fit to the neutron diffraction profile of \pr\ at 4 K.}
\label{fig3}
\end{center}
\end{figure*}
\\Our neutron powder diffraction measurements  were made using D20 at the ILL, Grenoble \cite{d20}, and showed evidence magnetic order below 230 K.  All reflections could be indexed with the primitive unit cell, i.e. $k\  =\  (0,0,0)$. The peaks showed an unusual temperature dependence (Fig. 3a), with several reaching a peak at 100 K with more complex behavior on further cooling. Above 100 K, the data could be accounted for using ordered Mn spins only (region I in Fig. 3b). Representational analysis of the possible magnetic structures allows AFM solutions with moments along all three axes. However, only a $z$ component was necessary to fit the data. The magnetic structure corresponds to nearest neighbour AFM coupling in the plane and ferromagnetic stacking of layers (i.e. C-type order). The refined moment reached a peak of $\sim$3\ \MuB \ at 100 K, which compares to a value of $\sim$ 5 \MuB \ expected for spin-only Mn$^{2+}$. Notably, this relative reduction ($\sim$40 \%) in moment from that expected in a local moment picture, is much less than that seen in iron pnictides such as LaFeAsO ($\mu$/$\mu_{exp}$ \  $\sim$0.1). In the region of the broad anomaly seen in the specific heat (80 K), the decrease in intensity of the principle magnetic reflections is accounted for by a flop of the Mn spins into the $ab$ plane (region II in Fig. 3b). We also detected a small moment attributed to induced ordering of the Pr$^{3+}$ sublattice, which shows a characteristic concave temperature dependence. The induced magnetic structure consists of antiferromagnetic bi-layers of Pr$^{3+}$ moments with the moments also lying in the $ab$-plane. This interplay between sublattices is similar to what was reported \ \cite{prfeaso} \ in PrFeAsO, except it is more obviously manifested due to the larger moments.
\\Below the T-O structural phase transition, we detected several new magnetic reflections which also indexed on the orthorhombic cell. The magnetic order in the low temperature phase is identical to that seen in region II, except the orthorhombic distortion allows us to distinguish the direction of the moments in the $ab$-plane (along the long $a$-axis). The saturated moments were 3.69(3) \MuB \  for Mn and 2.96(3) \MuB \  for Pr. The magnetic structure at 4 K is shown in Fig. 3d and the low angle part of the Rietveld fit to the 4 K data is shown in Fig. 3e.
\begin{figure}[tb!]
\begin{center}
\includegraphics[scale=0.3]{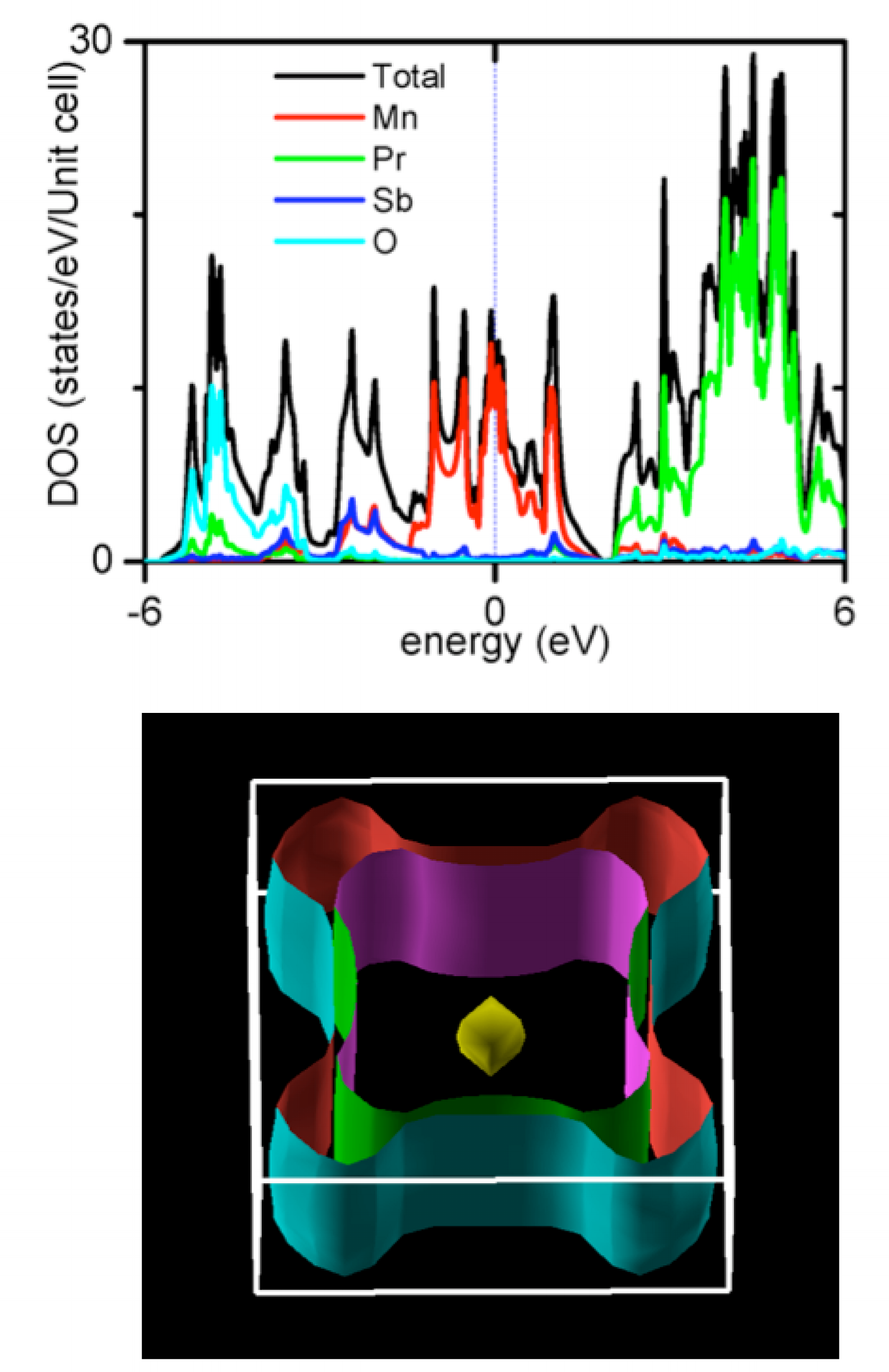}
\caption{(color online) (top) Total density of states for $P4/nmm$ \pr\ showing the large contribution of Mn 3$d$ states to the DOS around E$_{F}$. (bottom) Fermi surface of \pr\ in the first Brillouin zone, in contrast to the FeAs superconductors, no major nesting is observed.}
\label{fig4}
\end{center}
\end{figure}
\\The origin of the structural phase transition in \pr\ is clearly different from that found in the iron pnictides, which is driven by transition metal magnetism. Furthermore, related materials such as NdCoAsO, which also show interplay between both magnetic sublattices, do not show any distortion from tetragonal symmetry\ \cite{jan,ndco}. The magnetic order in the orthorhombic phase is also not obviously symmetry breaking as it is similar to that found in region II. Finally, C-type magnetic order implies that $J_{1} \ >> \ J_{2}$ in the MnSb planes, i.e magnetic frustration is negligible. What then is the origin of the structural phase transition? As Pr$^{3+}$ has a $4f_{2}$ configuration, one possibility is multipolar order. This scenario would result in a subtle crystallographic distortion of the Pr site, which we detect with our diffraction measurements (Fig. 3c). Below 35 K we find that the Pr-O distance splits into two and the Pr-Sb distance contracts. For comparative purposes we also plot the maximum variation found in these distances in PrFeAsO below T$_{N}$(Pr). These results are thus consistent with local $f$-electron degrees of freedom driving the structural transition. Ferro-multipolar order with $k \ = \ (0,0,0)$, would be compatible with the observed magnetic structure, however our measurements do not directly probe this order parameter. Confirmatory investigations with resonant X-ray scattering are clearly justified.
\\Experimentally, the magnetism in \pr \ appears to involve large local moments, which barely couple to the conduction electrons.
To explore further the interplay between structure and magnetism in
\pr\, we performed density functional calculations using the
Perdew-Burke-Ernzerhof generalized gradient approximation (GGA) in
the WIEN2k code \cite{blaha} based on the room temperature structure
where the open core approximation is employed for the Pr $4f$ states. Our
major findings are shown in Fig. 4. Similarly to the
iron pnictides, we find that the density of state (DOS) at the
Fermi level is almost entirely made up of transition metal
$d$-states. Among these we find that the d$_{x2-y2}$ and
d$_{xz}$/d$_{yz}$ states are most prominent where the local coordinates
are $x=a$, $y=b$, and $z=c$. The
calculated Fermi surface of \pr\ is also shown in Fig. 4, which
unlike the iron pnictides does not show prominent nesting. Instead
we find a rather three dimensional pocket at the zone centre and
warped sheets at the zone boundary. We also calculated the
non-interacting susceptibility and did not detect strong peaks at $q
\ = \ (\pi,\pi,0)$ or at $q \ = \ (0,0,0)$. These findings, together
with the experimental observations above, show that the magnetism in
\pr \ is a result of local exchange interactions rather than Fermi
surface nesting, despite the good electrical conductivity.
\\Furthermore, we performed both spin polarized GGA (sp-GGA) and local spin
density approximation (LSDA) calculations based on the low temperature
orthorhombic structure. Without considering the magnetization on Pr, the
C-type Mn magnetically ordered state is obtained with calculated
magnetic moments of  $\mu_{z}$ = 3.59~$\mu_B$ (sp-GGA) and
3.46~$\mu_B$ (LSDA), which are comparable to experimental
observation, indicating that spin fluctuations in \pr\ are not as
strong as in iron pnictides. However, a metallic ground state is only realised if the moment size is constrained to 3 $\mu_{B}$. We additionally found that introducing AFM Pr order destabilises the Mn order indicating strong coupling between sublattices as experimentally observed. However, further investigations have to be made to
understand the microscopic origin of the spin-flop process and the origins of the structural distortion at 35 K.
\\In conclusion, we have
reported the first example of a structural distortion driven by
magnetism in the ZrCuSiAs family of compounds outside the
ferro-pnictides. This transition is not driven by Fermi surface nesting, and despite the large moments, magnetic frustration is apparently unimportant. Instead the distortion in \pr \ is driven by
$f$-electron degrees of freedom. Future investigations using X-ray or neutron scattering will help to clarify this transition and determine any relation to those found in the iron pnictides.

We acknowledge the Helmholtz Zentrum Berlin for funding and the European Synchrotron Radiation Facility and Institute Max von Laue-Paul Langevin for access to their instruments. We thank K. Proke\v{s}, D. Le and T. Chatterji for useful discussions. RV, RP and DNA thank the Deutsche Forschungsgemeinschaft for financial support under SPP 1458.


\end{document}